\documentstyle[twocolumn,aps,prl,overcite]{revtex}

\tighten
\begin{document}
%\draft

\title{Quasi-Unit Cell Model for an Al-Ni-Co Ideal Quasicrystal
based on Clusters with Broken 10-fold Symmetry}

\author{Eiji Abe$^1$, Koh Saitoh$^2$, H. Takakura$^1$, 
A.P. Tsai$^1$, P. J. Steinhardt$^3$, and H.-C. Jeong$^4$ \\}
\address{$^1$National Research Institute for Metals, 
1-2-1 Sengen, Tsukuba 305-0047, Japan}
\address{$^2$Research Institute for Scientific Measurements,
Tohoku University, 2-1-1 Katahira, Sendai 980-8577, Japan}
\address{$^3$Department of Physics, Princeton  University, 
Princeton, NJ 08544, USA}
\address{$^4$Department of Physics, Sejong University, 
Kwangjinku, Seoul 143-747, Korea}
\date{\today}

\maketitle
\begin{abstract}
We present new evidence supporting the quasi-unit cell
description of the $Al_{72}Ni_{20}Co_{8}$ decagonal quasicrystal
which shows that the solid is composed of repeating,
overlapping decagonal cluster columns with broken 10-fold
symmetry.  We propose an atomic model which gives 
a significantly improved fit to electron microscopy experiments 
compared to a previous proposal by us and to alternative proposals 
with10-fold symmetric clusters.
\end{abstract}% 
%%ENDCHANGE
\pacs{61.44.Br, 61.66.Dk, 61.16.Bg}

The $Al_{72}Ni_{20}Co_{8}$ decagonal phase is one of the 
best-characterized quasicrystalline materials\cite{1} and an 
excellent candidate for comparing structural models of 
quasicrystals. This ideal, highly perfect
quasicrystal is reproducible as a single phase in the 
$Al_{72}Ni_{20}Co_{8}$ alloy 
annealed at 1170K followed by water-quenching.  

In a recent paper\cite{2} (henceforth referred to as Paper I),
we presented an array of evidence
that the atomic structure 
of $Al_{72}Ni_{20}Co_{8}$  conforms to the quasi-unit 
cell picture.\cite{3} In this  picture, the atomic structure can be 
decomposed into a single, repeating cluster (the quasi-unit cell)
which shares atoms with neighbor 
clusters according to specific overlap rules.
As first shown by P. Gummelt,\cite{4}
overlap rules can be sufficient to
insure a unique structure that has perfect
quasiperiodic translational order and the same 
symmetry as the Penrose tile picture based on two repeating 
tiles with edge-matching rules.\cite{5}
However, by reducing the structure to only one repeating unit,
 the quasi-unit cell picture leads to a simple 
description of the atomic structure\cite{2,3,4,6} 
and requires only simple 
energetics\cite{2,3} to explain why quasicrystals form and how 
they grow.
Our study showed that, based on high-angle annular 
dark-field scanning
transmission electron microscopy (HAADF-STEM) imaging\cite{7} 
which highlights the transition metal (TM) sites 
(Z-contrast\cite{8}), the quasi-unit cell for $Al_{72}Ni_{20}Co_{8}$ 
can be constructed from a 2~nm diameter, 
decagonal cluster column whose atomic arrangement
breaks 10-fold symmetry. We further proposed a specific atomic 
decoration which satisfy the observed centrosymmetric
$P10_5/mmc$ symmetry and 
the measured density and stoichiometry.  
An interesting feature is that the resulting atomic 
structure  cannot be described as a simple decoration of 
tiles in a Penrose tiling even though it can be described 
as a ``covering" composed
of identical,  overlapping,  decagonal cluster columns.\cite{2}

In this paper, we present an improved atomic model 
(Fig. 1(a)) and new experimental evidence
for a key feature arising from the quasi-unit cell picture: namely, 
the atomic decoration of the quasi-unit cells has broken
10-fold symmetry isomorphic to the broken symmetry of the 
overlap rules.
The improved model is motivated  by 
recent criticisms by Yan and Pennycook (YP) who presented 
a very high resolution HAADF-STEM image that revealed
some disagreements with our earlier atomic description.\cite{9}
YP advocate an alternative structure which conforms to the
quasi-unit cell picture in that it is composed of a single, 
decagonal cluster column obeying the same overlap rules. 
However, the YP proposal for the  atomic decoration of the 
cluster (Fig. 1(b)) exhibits perfect 10-fold symmetry, similar 
to  some previous structural models of 
$Al$-$Ni(Cu)$-$Co$ quasicrystals.\cite{10,11,12,13}

The central issue highlighted by YP is whether the decagonal 
clusters have ``intrinsic" broken symmetry, as we suggest, 
or whether the fundamental atom cluster is 10-fold symmetric 
and the broken symmetry is only  a consequence of random 
chemical and occupational (vacancy) disorder, as YP suggest.\cite{9}
(We use the term ``intrinsic" to refer to  broken
symmetry that is a built-in aspect of the atomic decoration, as 
opposed to a random disorder effect superposed on a fundamental 
cluster with 10-fold symmetry.) Although both possibilities are 
consistent with the quasi-unit cell picture,
the question is significant because we have argued that
the broken symmetry is a direct manifestation of the overlap 
rules or energetics of the ground state. According to this point 
of view, the  appearance of the broken symmetry clusters in images 
is fundamental to the structure and  a sign that the sample is 
nearly ideal. In contrast, YP conclude that any broken symmetry is 
accidental, due to random chemical and occupational disorder.

Our first step is to address the criticisms of our earlier 
model by YP based on HAADF-STEM images.  We modify the 
atomic decoration of the quasi-unit cell; see Fig. 1(a). 
The changes from the previous model are: (1) switch  $Co$ atoms 
from the interior of  kite-shaped regions (near the acute corner) 
with $Al$ on the edges  of the  kite-shaped region in Fig. 1(a);
and (2) add three $Al$ atoms in the central kite interior.
As in our earlier model, the quasi-unit cell has an atomic 
decoration that breaks 10-fold symmetry in a pattern isomorphic 
to the configuration of kites inscribed in the decagon, 
which represent the overlap rules.\cite{2,3} 
(The ``kite" is the convex polygon inscribed with light-blue 
 in the decagon of Fig. 1(a), which is a mnemonic for representing 
the overlap rules; neighbor decagons can overlap only if any 
kite-regions in the overlap region lay precisely on top of one another.)
Also, as before, the atomic 
model cannot be interpreted as  a simple Penrose 
tile decoration. The improved model now fits all TEM imaging data
(Figs. 2-4), reproduces the 
observed $P10_5/mmc$ symmetry, and 
has density (3.98 g/cm$^{3}$) and stoichiometry 
$Al_{71.2}Ni_{20.5}Co_{8.2}$ consistent with the measured 
values\cite{2} (3.94 g/cm$^{3}$ and $Al_{72}Ni_{20}Co_{8}$, with 
uncertainty $< 2$\%).  

\begin{figure}
%\epsfxsize=3.2in
%\epsfbox{abef1rev.eps}

\caption{Two competing models for the atomic decoration 
of the decagonal  (2~nm) quasi-unit cell for 
$Al_{72}Ni_{20}Co_{8}$: (a) our proposed model with broken 
10-fold symmetry and (b) an alternative model with unbroken 
10-fold symmetry but with accidental symmetry-breaking 
introduced in the central region due to chemical and 
occupational (vacancy) disordering.
Solid circles represent level  $c=0$ and open circles
represent $c=1/2$ along periodic $c$-axis. Figure (a) also 
includes atoms added by overlap 
of neighbor 
clusters (encircled in black).} 
\label{}
\end{figure}

Fig. 1(b) shows a competing, 
10-fold symmetric decoration, similar to the suggestion
by YP.\cite{9} 
The 10-fold symmetry may not be apparent because 
YP must introduce chemical and occupational (vacancy)
disorder in the central ring of atoms, typically 5 TM and 5 $Al$ sites 
shown in Fig.1(b), in order to produce an acceptable fit to the 
HAADF-STEM imaging. The result produces TM cluster pairs 
similar to Fig. 1(a). Nevertheless, it is important to note for the 
present high-resolution transmission electron microscopy (HRTEM) 
study that the 
configuration of atomic positions --- ignoring whether they are 
occupied by $Al$ or TM --- is 10-fold symmetric as proposed 
by YP and in earlier models.\cite{8,9,10,11} 

Another feature of Fig. 1(b) is that 
most sites are ``mixed," having fractional occupancy by $Al$, TM 
or vacancy depending on random  disorder.
Mixed and statistically occupied atomic sites are 
required for the 10-fold symmetric model to obtain a reasonable 
stoichiometry and density (assuming fully occupied sites,
the  atomic density, approximately 
0.073\AA$^{-3}$,\cite{10} is  too high for a 
densely-packed metallic structure, corresponding to 4.35 g/cm$^3$ if
one assumes the observed stoichiometry).
In our model, 
Fig. 1(a), each site is purely $Al$, $Ni$ or $Co$. 
It is possible for a vacancy in one decagon to be a filled site in
another decagon due to additional atoms contributed by 
overlap of neighbors in one case and not the other.
This effect is not random, though; the distribution of filled
(or unfilled) sites is quasiperiodically correlated in the ideal limit.
In this sense, 
both Figs.1(a) and (b) are only representative; depending on 
neighbors. The configuration inside the decagon in our model 
can have three (minor) variations,\cite{14} and many more variations 
are possible in the chemical and occupational disorder model.
Here, though, we have intentionally shown  examples of 
clusters from each model
that are most nearly the same to show how even they 
can be distinguished. 

We now consider the experimental evidence
for  intrinsic broken symmetry (Fig. 1(a)) based on a combination 
of  HAADF-STEM imaging (highlights the TM sites) and HRTEM 
(total projected potential).

First, in our analysis of a large HAADF-STEM image in Paper I, 
we found that essentially all  clusters exhibit the same broken 
10-fold symmetry (a triangle of TM spots, as shown in Fig. 2(c))  
in the center-most region of each cluster. Some spots are 
elongated, suggesting closely-spaced,   column
pairs of TM atoms, as occurs in both models in Fig. 1.
The triangle of spots breaks the symmetry within a decagon
in a manner isomorphic to the overlap rules, which are 
represented by kite-shaped decorations inside each decagon;  
see Fig. 1(a). 
YP propose to  produce a similar triangle through chemical 
disorder, as shown in Fig. 1(b). Even if one accepts the notion 
that chemical disorder would somehow produce a similar
triangular pattern in nearly  all clusters, the problem arises 
that the orientation of the triangle of spots is correlated 
across the image in accord with the overlap rules. 
This correlation is apparent in the HAADF-STEM 
figure in Paper I which shows that every kite-shaped region
in the overlap pattern of decagons corresponds to a triangle 
of spots with matching orientation.
The orientational correlation suggests strongly a chemical 
ordering between the $Al$ and TM atoms
and is inconsistent with random chemical disorder.

A second problem with the 
chemical and occupational disorder proposal is that it is 
difficult to explain why highly perfect samples of $AlNiCo$ 
only occur for a narrow composition range. The highly 
perfect $AlNiCo$ phase appears to reveal diffuse streaks 
or superlattice reflections when its composition 
deviates from the $Al_{72}Ni_{20}Co_{8}$ stoichiometry 
by only a few (atomic) percent for all constituent 
elements.\cite{1} Note that the perfection of the sample 
is even sensitive to the ratio of $Ni$ to $Co$, suggesting
that the two TM atoms, also, may lie at specific positions 
in the structure as proposed in our model (a variant $Ni/Co$
ordering could also be considered, as discussed later). 
If the structure could tolerate sufficient disorder (a large 
number of mixed sites in Fig.1(b)) to transform all or nearly all 
decagonal clusters to 10-fold symmetry breaking clusters, 
then one would expect that the single phase region would 
extend to a wider composition range than a order
of a few atomic percent, but it does not.

The above two discussions lead to a conclusion
that  $Al_{72}Ni_{20}Co_{8}$
is a quasiperiodic intermetallic compound with nearly
perfect atomic order and close to its ideal stoichiometry.
Some alternative models presume that the structure must 
be significantly disordered because the phase is 
thermodynamically stable only at 
high-temperature.\cite{9,10,11,12,13}
However, we would suggest that the
situation need not be  different from the case of crystalline 
phases which are thermodynamically stable only at
high-temperature.
In either case, that state of lowest free energy (at finite
temperature) can be described in terms of an ideal ordered state
even though, upon quenching, there exist some defects and disorder,
as discussed later.

\begin{figure}
%\epsfbox{abef2.eps}
\caption{(a) HRTEM image of the high quality sample of 
$Al_{72}Ni_{20}Co_{8}$, taken at near-edge of a cleavage 
grain (Fourier-filtering was made to subtract the 
background).
(b) A decagonal region showing rectangular region  used for
comparing real versus predicted image contrasts (see Fig. 3). 
(c) The model in Fig. 1(a) superimposed on HAADF-STEM
image, confirming a validity of the TM sites.}
\label{1}
\end{figure}

A third argument for intrinsic broken symmetry is based on new,
direct evidence from HRTEM by a 400kV TEM with a resolution 
of 0.17nm. 
For HRTEM observation, samples were crushed and then dispersed
on perforated carbon films supported on Cu grids. 
Simulation of image contrasts was performed using 
MacTempas program (Total Resolution, Inc.) 
whose calculation is based on the multislice method.

Fig. 2(a) shows the HRTEM structure image of the 
$Al_{72}Ni_{20}Co_{8}$
taken from the tenfold symmetry axis under nearly the 
Scherzer defocus ($-45$~nm for the present microscope),
which reflects the projected electrostatic potential: 
dark regions correspond to the projected atomic 
positions.\cite{10} Some 2~nm decagonal 
clusters have been  outlined in the image to guide the eye. 
Viewing carefully the interior  of the decagon
(rectangle-region of Fig. 2(b)), one notices that the contrasts
appear to break 10-fold symmetry. 
Similar symmetry breaking is found in each of the 
decagon regions in Fig. 2(a).

\begin{figure}
%\epsfbox{abef3.eps}
\caption{Simulated images of the
atomic models in Fig. 1  are shown in (a) and (b), respectively.
Both were calculated with $-45$~nm defocus and 3.6~nm 
thickness. Differences between the observed and simulated 
image contrasts of the rectangle regions are shown in (c) and (d).}
\label{2}
\end{figure}

These images should be compared to Figs. 3(a) and (b), which 
show the calculated HRTEM image contrasts for the two 
models in Fig. 1. Note that the decagon center 
in  Fig. 3(a) exhibits a triangle feature, 
(10-fold symmetry-breaking) while  Fig. 3(b) reveals a nearly 
perfect circle.  The introduction of chemical disorder 
in the central decagonal ring in Fig. 1(b)
to match the HAADF-STEM image has not significantly 
affected the HRTEM image. The observed  image in Fig. 2(a) is 
more similar to the broken symmetry model in Fig. 3(a).
To confirm this, we have computed the
difference between the observed and calculated images
over the rectangular region outlined in Figs. 2(b) and 3(a) 
and 3(b); the residual intensities are  shown in Figs. 3(c) 
and (d). Clearly, the model with intrinsic broken symmetry 
fits better. We emphasize that the difference originates
from symmetry breaking in the arrangement of atomic 
sites (regardless of $Al$ or TM), such as a slight 
displacement (0.95{\AA} shift) of 3 $Al$ atoms at the 
core\cite{16} (see arrows in atomic models in Fig. 3(c)) 
that breaks the symmetry of the decagonal ring in Fig. 3(d).
Note that the image contrasts at the core will appear 
to be more symmetric  if there are phason defects which flip
decagons on some layers but not others within the thickness 
observed.\cite{Widom}  
The more symmetric "b" clusters in the HAADF images
shown by YP,\cite{9}
which are relatively rare compared to the 10-fold symmetry
breaking clusters, may be explained by this effect.

\begin{figure}
%\epsfbox{abef4rev.eps}

\caption{Observed and computed (with 3.6nm thickness)
through-focus HRTEM images at different defocus values.}
\label{3}
\end{figure}

We have also performed a through-focus HRTEM study. 
Since the phase difference between direct and diffracted 
beams is altered by changing the focus value in HRTEM, the 
image contrasts appear to be different depending on the 
focus values.\protect{\cite{17}} (a low-pass filter effect
to select a desired diffracted beam to be 
the major contribution for imaging.)  The systematic change 
in the image contrasts with changing focus
must be reproduced by the atomic model.
The top row of Fig. 4 shows HRTEM
images taken at smaller ($\Delta f=-25$~nm)
and larger  ($\Delta f=-70$~nm) defocus values 
than the Scherzer value, -45~nm.  
The corresponding calculated images are shown 
for the broken symmetry model (Fig.1(a)) in the second row 
and for the 10-fold symmetric model (Fig.1(b)) in the third row.
Clearly, the  observed
defocused images exhibit the 10-fold symmetry breaking 
contrasts, confirming
the result with Sherzer defocus (Fig. 3) and supporting
models  intrinsic broken 10-fold symmetry. 

Recently, M. Widom and coworkers\cite{Widom} have completed
a total energy based prediction of the structure of $AlNiCo$, 
in which 2nm decagonal clusters with broken 10-fold symmetry 
emerge in the lowest energy configuration with
nearly identical assignments of $Al$ and $TM$ positions as shown Fig.1(a).
(The total energy, they find,  prefers $Ni-Ni$  pairs to our $Ni-Co$ pairs;
switching the $Co$ with isolated $Ni$ in Fig.1(a)
produces a variant model that also gives a reasonable fit.)
The convergence between our experimental test of the 
quasi-unit cell picture and the recent
theoretical analysis suggests that there is real hope for 
obtaining a definitive atomic model for $AlNiCo$ from 
a further structural refinement (by diffraction method)
on the present ideal model. This implies that a theoretical 
understanding of the structure may guide the discovery 
of new quasicrystals.

This research was supported by CREST of Japan Science 
and Technology Corporation, US Department of Energy grant 
DE-FG02-91ER40671 (Princeton) and by the  Korea 
Research Foundation Grant KRF-IDR98-B0001.


% \begin{table}
% \caption{}
% \label{}
% \end{tabular}
% \end{table}

\begin{references}
\bibitem{14} Depending on the symmetry and arrangement 
of atoms in the quasi-unit cell, there can be up to nine 
variations of arrangements in the interior of a decagonal 
cluster when overlap between neighbor clusters in included. 
In our proposed model, there are only three variations.
\end{references}
\end{document}